\documentclass[hidelinks,12pt]{article}
\pdfoutput=1

\usepackage[utf8]{inputenc} %
\usepackage[toc,page]{appendix} %
\usepackage[explicit]{titlesec} %
\usepackage{hyperref}
\usepackage[makeroom]{cancel}
\usepackage{xcolor}
\usepackage[parfill]{parskip}
\usepackage[affil-it]{authblk}
\usepackage{fancyhdr}
\usepackage{xparse}
\usepackage{geometry}

\usepackage{mathtools}
\usepackage{amssymb}
\usepackage{amsthm}
\usepackage{amsfonts}
\usepackage{verbatim}
\usepackage{enumerate}
\usepackage{graphicx}
\usepackage{accents}
\usepackage{bbm}
\usepackage{bm}
\usepackage{mathrsfs}
\usepackage{xfrac}
\usepackage{lmodern} %
\usepackage{fix-cm} %

\usepackage{physics}
\usepackage{tensor}
\usepackage{siunitx}
\usepackage{wasysym}
\usepackage{slashed}

\usepackage{graphicx}
\usepackage{caption}
\usepackage{subcaption}
\usepackage{epstopdf}
\usepackage{pgfplots}

\usepackage{tikz} %
\usetikzlibrary{fadings}
\usetikzlibrary{arrows,shapes,positioning}
\usetikzlibrary{decorations.markings}
\usepackage[backend=biber, sorting=none]{biblatex}

\usepackage{todonotes}

\usepackage{listings}

\numberwithin{equation}{section} 

\pgfplotsset{compat=1.14}

\makeatletter
\renewcommand{\maketitle}
{\begingroup
	\begin{flushright}
		{\large SCIPP 18/02\\ \large ACFI-T17-08}
	\end{flushright}

	\vskip 1.2cm

	\begin{center}
		{\LARGE\bf \@title}

		\vskip 1.4cm

		{\@author}

		\vspace{0.3cm}
	\end{center}

	\vskip 4pt
\endgroup}
\makeatother

\renewcommand{\r}{\scriptr}

\renewcommand{\v}[1]{\vec{#1}}

\renewcommand{\b}[1]{{\bf #1}}

\newcommand{\cA}{\mathcal{A}}

\newcommand{\cL}{\mathcal{L}}
\newcommand{\cM}{\mathcal{M}}

\newcommand{\dbar}{{\mathchar'26\mkern-12mu \mathrm{d}}} %
\newcommand{\ddbar}{}
\DeclareDocumentCommand\ddbar{ o g d() }{ %
	\IfNoValueTF{#2}{
		\IfNoValueTF{#3}
			{\dbar\IfNoValueTF{#1}{}{^{#1}}}
			{\mathinner{\dbar\IfNoValueTF{#1}{}{^{#1}}\argopen(#3\argclose)}}
		}
		{\mathinner{\dbar\IfNoValueTF{#1}{}{^{#1}}#2} \IfNoValueTF{#3}{}{(#3)}}
	}
\newcommand{\ddelta}[2][1]{ %
	\if#11
	\delta \left( #2 \right)
	\else
	\delta^{\left( #1 \right)} \! \left( #2 \right)
	\fi
}
\newcommand{\func}[3][1]{ %
	\if#11
	\mathinner{#2 \! \left( #3 \right)}
	\else
	\mathinner{#2^{#1} \! \left( #3 \right)}
	\fi
}
\newcommand{\dfunc}[3][0]{ %
	\if#10
	\mathinner{#2 \! \left( #3 \right)}
	\else
	\mathinner{#2^{\left(#1\right)} \! \left( #3 \right)}
	\fi	
}
\newcommand{\Func}[3][1]{ %
	\if#11
	\mathinner{#2 \! \left[ #3 \right]}
	\else
	\mathinner{#2^{#1} \! \left[ #3 \right]}
	\fi
}
\newcommand{\dFunc}[3][0]{ %
	\if#10
	\mathinner{#2 \! \left[ #3 \right]}
	\else
	\mathinner{#2^{\left(#1\right)} \! \left[ #3 \right]}
	\fi
}

\newcommand{\littleo}[1]{\func{o}{#1}} %

\newcommand{\gammafunc}[1]{\mathinner{\Gamma \! \left( #1 \right)}}

\newcommand{\mplanck}{{m_{\mathrm{P}}}} %

\hypersetup{
	pdftitle={Behavior of Cross Sections for Large Numbers of Particles},
	pdfauthor={Michael Dine, Hiren H. Patel, Jaryd F. Ulbricht}
}

\renewcommand{\vec}[1]{\vb{#1}}  

\def\beq{\begin{eqnarray}}
\def\eeq{\end{eqnarray}}
\def\bea{\begin{eqnarray*}}
\def\eea{\end{eqnarray*}}

\title{Behavior of Cross Sections for Large Numbers of Particles}

\author{Michael Dine}
\author{Hiren H. Patel}
\author{Jaryd F. Ulbricht}

\affil{Santa Cruz Institute for Particle Physics and\\ Department of Physics, University of California at Santa Cruz\\ Santa Cruz CA 95064}

\begin{document}
\begin{titlepage}
\maketitle

\begin{abstract}
It has been suggested that scattering cross sections at very high energies for producing large numbers of Higgs particles may exhibit factorial growth, and that curing this growth might be relevant to other questions in the Standard Model. We point out, first, that the question is inherently non-perturbative; low orders in the formal perturbative expansion do not give a good approximation to the scattering amplitude for sufficiently large \(N\) for any fixed, small value of the coupling. Focusing on \(\lambda \phi^4\) theory, we argue that there \textit{may} be a systematic approximation scheme for processes where \(N\) particles near threshold scatter to produce \(N\) particles, and discuss the leading contributions to the scattering amplitude and cross sections in this limit. Scattering amplitudes do not grow as rapidly as in perturbation theory. Additionally, partial and total cross sections do not show factorial growth. In the case of cross sections for \(2 \to N\) particles, there is no systematic large \(N\) approximation available. That said, we provide evidence that non-perturbatively, there is no factorial growth in partial or total cross sections.
\end{abstract}

\end{titlepage}

\section{Introduction}

The perturbation expansion of Greens functions in quantum field theories is typically asymptotic, with the coefficient of the \(n\)'th order terms exhibiting factorial growth in \(n\)\cite{largeorderreview}. This can be attributed to the factorial growth in the number of Feynman diagrams with \(n\).%
For some time it has been noted that there is similar growth in the amplitudes for processes in scalar field theories with large numbers of final state particles \(N\), e.g. in \(2 \to N\) processes, already at the level of the leading order Feynman diagrams\cite{goldberg,rubakovson,sonmultiparticle,brownmultiparticle}. This happens because, near threshold, the amplitudes are only very weak functions of momenta, and there are \(N!\) ways of rearranging the various final state particles in the lowest order diagrams. At extremely high energies estimates are more challenging, but there would seem likely to be factorially large numbers of contributions to amplitudes with the same sign. Various attempts have been made to compute or estimate the behavior of amplitudes and cross sections in the limit of large \(N\). These analyses are in some instances perturbative, and in some instances attempt to include non-perturbative effects. More recently it has been argued that the growth of amplitudes implies a physical cutoff at energies much less than the Planck mass \(\mplanck\), thereby reducing the severity of the hierarchy problem\cite{higgsplosion1,higgsplosion2}.

In this paper, we take a critical look at the question of the growth of scattering amplitudes and cross sections at very large \(N\) in \(\lambda \phi^4\) theory. We start with the simple observation that the problem is \textit{inherently non-perturbative} for \(N \gg 1/\lambda\): at each order in \(\lambda\), one obtains additional powers of \(N\); the expansion parameter is \(\lambda N\). We will focus on two classes of processes: \(2 \to N\) scattering and \(N \to N\) particle scattering. We will work near threshold (with \(\abs{\vec{p}} = \epsilon m\), where \(m\) is the particle mass and \(\epsilon\) is a small number which \textit{does not} scale with \(N\)). In the first case, the scattering amplitude in lowest order of perturbation theory grows as \(N!\). Bose symmetry gives a \(1/N!\) factor, and the phase space integral gives neither \(N!\) enhancement or suppression. So one has a cross section which grows as \(N!\). In the case of \(N \to N\) scattering there are, at large \(N\), of order \(\qty(2N)!\) independent contributions to the scattering amplitude at low orders, suggesting \((2N)!\) growth of the scattering rate. But even near threshold the amplitudes have complicated dependence on the momenta. If we focus on the most singular momentum region, the perturbative rates do not exhibit factorial growth, and we give a crude argument that this singular region dominates the cross section.

In either case the perturbative analysis is not reliable when \(N \gtrsim \lambda^{-1}\), and we would like some insight (and ideally a systematic approximation scheme) into these processes. In the case of \(N \to N\) scattering we describe a non-perturbative computation which we suspect to be the leading approximation in a systematic expansion in \(1/\sqrt{N}\). This yields a cross section which \textit{decreases} factorially with \(N\) as \(N\) becomes very large. In the case of \(2 \to N\), we adopt a different approach.   There does not appear to be a simple reorganization of the problem which would permit a systematic approximation in \(1/N\). We describe an admittedly crude computation which suggests, again, that amplitudes grow more slowly than in perturbation theory and cross sections don't exhibit factorial growth.

As a strategy to explore the non-perturbative behavior we consider the problem from the perspective of the path integral. To cast the scattering problem in this language we employ the LSZ formalism. To render the LSZ result in a fashion which is convenient for a path integral analysis we can proceed in two ways. One approach is to reorganize the Feynman diagram expansion into a finite (power of \(N\)) number of classes of diagrams, each of which can be expressed as an integral of a Green's function weighted by external wave functions. Each of these Green's functions, in turn, can be simply expressed as a path integral. In some cases, taking the large \(N\) limit allows one to evaluate these by semiclassical methods. This statement relies critically on the use of normalizable wave packets, and the fact that there is a small space-time region where all of the wave packets coincide. From the path integral perspective, it is only in this small region of space-time, involving only a small subspace of the field space, where the large value of \(N\) is important. The amplitudes do indeed appear to be dominated by a particular classical configuration; more precisely a particular region of integration. This allows a simple determination of the scaling with \(N\), and a possible systematic expansion in \(1/N\) for the case of \(N \to N\) scattering. For \(2 \to N\), we adopt a different approach. The analysis is not systematic but strongly suggestive. As in the \(N \rightarrow N\) case, this leads to an expectation that the growth of the amplitude is slower than in perturbation theory, and that the cross sections do not exhibit factorial growth with \(N\).

In the rest of this note we investigate these questions. Other critiques and responses have appeared elsewhere\cite{higgsplosioncritique1,higgsplosioncritique2,higgsplosionconsistency}; the issues we raise are somewhat different. We will set aside the question of whether the theory has a sensible limit as its ultraviolet cutoff is taken to \(\infty\), but will explain the analysis which leads to the results stated above. We first review, in  section \ref{sec:perturbative2ton}, the leading perturbative result for the cross section in \(2 \to N\) scattering. We illustrate features of \(N \to N\) scattering in section \ref{sec:nton}. While there are a vast number of diagrams we isolate a subset of them by examining a singular kinematic limit, where in low orders of perturbation theory there is not factorial growth of the coefficient of the most singular behavior. We then argue that while the non-singular diagrams are far more numerous, they are suppressed by factors of \(1/N\) relative to the singular diagrams.

In both cases, as we note in section \ref{sec:limitations}, perturbation theory becomes unreliable when \(N \gtrsim 1/\lambda\), so any would-be conflicts with unitarity should arguably be viewed with skepticism when derived from a perturbative framework. To obtain some insight into the non-perturbative problem, in section \ref{sec:onedintegral}, we review the behavior of a simple one-dimensional integral (zero-dimensional field theory) which possesses some of the features expected of actual \(\phi^4\) theory. In section \ref{sec:wavepacketscattering} we review some basic aspects of scattering of wave packets in non-relativistic quantum mechanics in order to set the stage for our discussion of some aspects of the LSZ formula in section \ref{sec:lsz}. In particular, we focus on the scattering amplitude for normalizable initial and final states as well as aspects of the path integral. In section \ref{sec:nonperturbativenton} we explain why, in the case of \(N \to N\), this reorganization of the path integral is particularly effective and why the approximation appears systematic. We also show that the amplitude is substantially reduced over the leading perturbative contribution. On the other hand, for \(2 \to N\), the approach provides, at best, a crude estimate. A more useful strategy, developed in section \ref{sec:nonperturbative2ton}, involves the study of an effective action for \(2 \to N\) processes, where the \(N\) final state particles are near threshold. This problem can also be expressed in path integral language and one can obtain a recursion relation for \(\Gamma_{2 \to N}\) for different values of \(N\). The recursion relation, for \(N \ll 1/\lambda\), reproduces the perturbative result, but it leads to slower growth at larger \(N\). This, in turn, translates into a cross section which does not exhibit factorial growth.

In our concluding section we remark on implications of this work for unitarity at large \(N\), and also suggest possible further directions which might give greater control over this particular limit of quantum field theories.

\section{Perturbative Analysis of \texorpdfstring{\(2 \to N\)}{2 to N} Scattering in \texorpdfstring{\(\lambda \phi^4\)}{Scalar Field} Theory}
\label{sec:perturbative2ton}

We begin this section by reviewing a conventional perturbative analysis~\cite{goldberg,rubakovson,sonmultiparticle,brownmultiparticle} of the production of \(N\) non-relativistic scalars near threshold in a \(\lambda \phi^4\) theory from an initial state of two very highly energetic particles. The case where \(N = 2 \times 3^{k}\) lends itself to a simple analysis. In that case there is a diagram where the two incident scattered particles produce two particles, and then each splits into three, each of these splits into three, and so on, \(k - 1\) times. All of the internal lines are far off shell, and one can neglect the small kinetic energies of the final states. As a result, there are \(N!\) nearly identical contributions to the amplitude.

\begin{figure}[ht]
  \centering
  \includegraphics[width=0.5\textwidth]{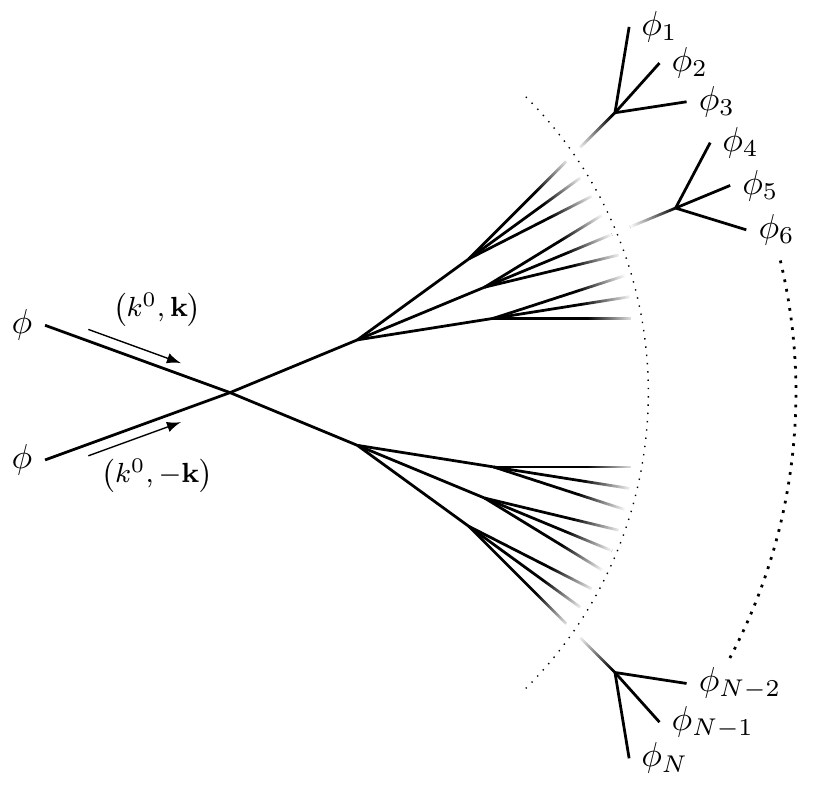}
  \caption{The initial state particles scatter to two very off shell particles, which then decay to three particles, and those particles decay to more particles, etc...}
  \label{fig:2ton}
\end{figure}

One might wonder whether there is suppression arising from the numerous propagators in the graph. The total number of vertices for this diagram is:

\begin{equation}
\begin{split}
V =& 1 + 2 \left( 1 + 3 + 3^2 + \dots + 3^{k -1} \right),\\[2mm]
=& 1 + 2 \left( \frac{3^{k} - 1}{2} \right),\\[2mm]
=& \frac{N}{2}.
\end{split}
\end{equation}

\noindent The propagator suppression can be determined in a similar fashion. Take the initial momenta to be 

\begin{equation}
\begin{split}
k_{1} \sim& \left( \frac{N m}{2}, 0, 0, \frac{N m}{2} - \Delta \right), \quad \left( N \to \infty \right)\\[2mm]
k_{2} \sim& \left( \frac{N m}{2}, 0, 0, - \frac{N m}{2} + \Delta \right), \quad \left( N \to \infty \right)
\end{split}
\end{equation}

\noindent Where \(\Delta \ll N m\). The propagator suppression is (in units with the meson mass, \(m\), set equal to one):

\begin{equation}
\func{h}{N} = \left[ \frac{1}{\left( \frac{N}{2} \right)^{2} - 1} \right]^{2} \times \left[ \frac{1}{\left( \frac{N}{2} \right)^{2} 3^{-2} - 1} \right]^{6} \times \ldots \times \left[ \frac{1}{\left( \frac{N}{2} \right)^{2} 3^{-2 \left( k - 1 \right)} - 1} \right]^{ 2 \times 3^{k - 1}} 
\end{equation}

\noindent Taking the log and neglecting the factors of \(- 1\) in the propagator gives:

\begin{equation}
\ln(\func{h}{N}) \sim - 3 \ln(3) \frac{N}{2} + 2 \ln(\frac{N}{2}) + \ln(3), \quad \left( N \to \infty \right)
\end{equation}

There is no net \(N!\) suppression from the large number of propagator factors. While we won't make claims as to the dominance of this set of diagrams, other classes of tree level graphs \textit{do} have an \(N!\) type kinematic suppression. For example, if most of the external lines connect in pairs to a single line, there is a substantial suppression. Considering only this class of diagrams, of which there are roughly \(\qty(3N/2)!\)\footnote{There are of order \(N!\) ways to rearrange the external lines and \(\qty(N/2)!\) ways to relabel the vertices. In the \(N \to \infty\) limit we write the product of \(N! \times \qty(N/2)! \Rightarrow \qty(3N/2)!\) The sense in which we mean this is that both sides of the \(\Rightarrow\) have factors of \(N^{cN}\) which are identical. In this paper we are primarily concerned with factors of \(N^{cN}\) and we will typically neglect factors such as \(a^{N}\) and \(N^{b}\).}, and after dividing by a factor of \(\qty(N/2)!\) coming from the \(N/2\) insertions of the interaction Lagrangian, we find:

\begin{equation}
\cM_{2 \to N} \sim 3^{- 3N/2} N! \left( \frac{\lambda}{4} \right)^{N} \quad \left( N \to \infty \right)
\end{equation}

In the cross section there is a factor of \(1/N!\) from Bose statistics. There is also a suppression from phase space when all of the particles are non-relativistic. If we assume that, in the center of mass frame, the total energy is 

\begin{equation}
\sqrt{s} = N \qty(1 + \epsilon) m,
\end{equation}

\noindent where we will think of \(\epsilon\) as small compared to \(1\) but not \(1/N\), we can consider final states where the momentum of each particle is of order \(\epsilon m\). Then the phase space factor is of order

\begin{equation}
\prod^{N}_{i = 0} \int^{\sqrt{\epsilon} m} \frac{\dd[3]{p_{i}}}{2 m} \sim \epsilon^{3/2 N} m^{2N}. \quad \left( \epsilon \to 0, N \to \infty \right) 
\end{equation}

\noindent Multiplying by the squared amplitude (restoring the factors of \(m\) and multiplying by the Bose statistics factor) the cross section goes like 

\begin{equation}
\sigma_{2 \to N} \sim \left( \frac{\epsilon}{3} \right)^{3/2 N} N! \qty(\frac{\lambda}{4})^{N/2}, \quad \left( \epsilon \to 0, N \to \infty \right).
\end{equation}

For \(\epsilon\) a small, but fixed, number the suppression from the phase space integral does not compensate the factorial growth in the amplitude for \(N > 1/\qty(\epsilon \lambda^{2})\). As we will elaborate in section \ref{sec:limitations}, the perturbative analysis is generally invalid once \(N \gtrsim \lambda^{-1}\). Our goal will be to get some idea of the behavior in this non-perturbative region. We will not be able to give a systematic analysis in \(\lambda\) and \(N\), but we will argue shortly that the non-perturbative growth of the amplitude is no faster than \(\qty(N/2)!\) As a result cross sections do not show factorial growth and there are no conflicts with unitarity. In the next section we will study a different class of processes exhibiting factorial growth in the number of Feynman diagrams, for which a systematic analysis \textit{may} be possible.

\section{\texorpdfstring{\(N \to N\)}{N to N} Scattering}
\label{sec:nton}

Another interesting class of processes involves \(N \to N\) scattering with all particles near threshold. Na\"{i}vely, given that there are of order \(\qty(N!)^2\) similar contributions to the amplitude, while the Bose statistics factor behaves as \(\qty(N!)^{-2}\), potentially leading to a rapid growth in the cross section. It is necessary, however, to consider possible kinematic enhancement and suppression.

In particular, in leading order in perturbation theory, there are kinematical enhancements of certain classes of diagrams. A particularly singular region occurs when all (spatial) momenta are non-relativistic, and pairs of momenta are nearly equal: \(\vec{p}_{i} = \vec{k}_{i+1} + \delta \vec{p}_{i}\). Then there are \(N!\) contributions where all of the internal lines are within \(\vec{p}_{i} \cdot \delta \vec{p}_{i}\) of the mass shell. To compute the amplitude we need to weight the Feynman diagrams with the initial and final wave functions and integrate. It is most convenient to work in momentum space.

\begin{figure}[ht]
  \centering
  \includegraphics[width=0.8\textwidth]{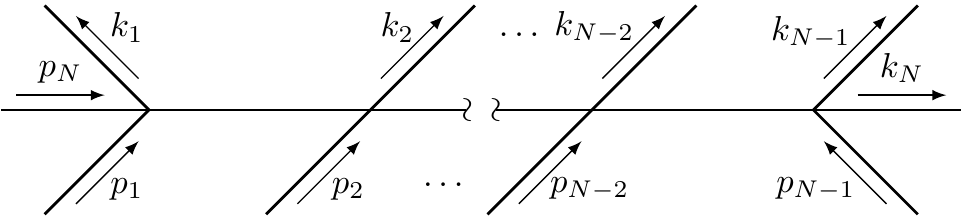}
  \caption{\(N \to N\) scatting. At every vertex an initial state particle scatters to a final state particle with almost no momentum exchange. All of the propagators are then nearly on shell.}
  \label{fig:nton}
\end{figure}

Consider the squared amplitude and integrate over final states for some fixed initial configuration. If the typical magnitude of the three-vector momentum is \(p\), then we restrict the final state momenta to all be of order \(p\), and focus on the integration region:

\begin{equation}
\vec{k}_{i} = \vec{p}_{i} + \delta \vec{k}_{i}, \quad i = 1, \ldots ,N; \quad \abs{\delta \vec{k}_{i}}
< \Delta k \ll p.
\end{equation}

\noindent In this region of phase space we focus on the \(\qty(2N)!\) diagrams where (at all but two vertices, involving \(\vec{k}_{N}\) and \(\vec{p}_{1}\)) the \(i^{\text{th}}\) final state particle emerges from the vertex from which the \(\qty(i + 1)^{\text{th}}\) initial state particle enters. The resulting amplitude behaves as

\begin{equation}
\cM_{N \to N} \sim \frac{\qty(2 N)!}{N!} \qty(\frac{\lambda}{4})^{N-1} \prod^{N-2}_{i=1} \frac{i}{2 \sum^{i}_{j=1} \qty(\vec{k}_{1} - \vec{k}_{j+1}) \cdot \delta \vec{p}_{j}}, \quad \qty(\delta \vec{p}_{i} \to 0, N \to \infty).
\end{equation}

\noindent Note that because of the pairing, there is an \(N!\) rather than \(\qty(2N)!\) factor. When we square the amplitude, integrate over the momenta \(\delta \vec{p}_{i}\), divide by the Bose statistic factors for the initial and final states, and restrict \(\abs{\delta \vec{p}_{i}} \ll p\) we obtain

\begin{equation}
\sigma_{N \to N} \sim \frac{p^{N} \qty(\Delta k)^{N}}{\mu^{2N}}, \quad \qty(N \to \infty). 
\end{equation}

\noindent With a similar restriction on the momenta, but without the pairing, one has a suppression by at least two powers of \(\Delta k / p\) for each unpaired momentum.

There are in fact, for fixed values of the initial and final momenta, vastly more diagrams which do not exhibit this pairing (\(\qty(2N)!\) rather than \(N!\)). However, these diagrams are further suppressed by powers of \(N\) coming from the large number of terms in some of the denominator factors. Roughly, the typical denominator is a sum of \(N^2\) terms, with random signs, so we might expect the sum to be of order \(N\) for roughly half the order \(N\) propagators. The full propagator suppression might then be of order \(1/N!\) from these diagrams, so that they are similar to those discussed  previously, without the extra kinematic enhancement. Moreover, the diagrams have phases, so we might expect a factor of order \(N!\) rather than \((2N)!\) contribution to the amplitudes from summing over all of the permutations in which not all  momenta are paired.

We don't claim this analysis to be more than suggestive, but we do believe it is plausible that with these kinematic restrictions, even at the lowest non-trivial order in perturbation theory, the factorial growth of the amplitudes is bounded from above\footnote{Here \(N\) is the total number of external particles. For a process that goes like \(N \to N\) there are \(2N\) external lines, so the bound should be stated as:

\begin{equation*}
\cM_{N \to N} = \littleo{N!} \quad \qty( N \to \infty).
\end{equation*}} by \(\qty(N/2)!\). This is despite the roughly \(\qty(3N/2)!\) Feynman diagrams that contribute to tree level processes. However, any estimate derived from perturbation theory is not reliable once \(N \gtrsim \lambda^{-1}\).

We will argue in the following sections that, non-perturbatively and in the region with small paired momenta, there is actually a factorial \textit{suppression} of the amplitude. We believe that this suppression can be rigorously established. For other regions of phase space, our arguments for suppression will be plausible but less rigorous. In any case, these considerations suggest that there is a region of phase space in which the cross section does not exhibit factorial growth, though it is enhanced by inverse powers of small momenta compared to na\"{i}ve expectations.

\section{Limitations of The Perturbative Analysis: Going Beyond}
\label{sec:limitations}
  
Perturbation theory is not reliable for either of the previously discussed processes for sufficiently large \(N\). If \(N \gg \lambda^{-1}\) then, for either \(2 \to N\) or \(N \to N\) scattering, one has an expansion in powers of \(\lambda N\). Once \(N \gg 1/\lambda\) any partial sum of the perturbative series becomes a very poor approximation of the true amplitude.  

A quick way to see that the expansion is a power series in \(\lambda N\) is to consider coupling constant renormalization. At leading order we have seen that the scattering amplitude behaves as \(\lambda^{N/2}\) or \(\lambda^N\) for \(2 \to N\) or \(N \to N\) processes respectively. The effect of one loop renormalization is to replace \(\lambda \to \lambda \qty(1 + A \lambda)\), so expanding in powers of \(\lambda\) we have a term of order \(A N \lambda\). This counting is easily seen to be general by examining other classes of Feynman diagrams.
 
We have established that, order by order in perturbation theory, as one approaches the relativistic limit scattering amplitudes exhibit factorial growth for processes such as \(2 \to N\) particles. However, it is not immediately clear how seriously to take these results. It is perhaps troubling to uncover a set of questions in quantum field theory which, even at weak coupling, we lack the tools to explore. We take some steps towards the non-perturbative study of large \(N\) amplitudes in the following sections.

\section{A Model for Factorial Growth of Amplitudes: A Simple One Dimensional Integral}
\label{sec:onedintegral}

Certain features of the large order behavior of perturbation theory in quantum mechanics and quantum field theory can be modeled by an ordinary integral. One of our goals in this paper will be to reduce the scattering amplitudes for large \(N\) to similar integrals. In a theory with \(\lambda \phi^4\) coupling, the one dimensional integral, 

\begin{equation}
\func{Z}{\lambda} = \frac{1}{\sqrt{2 \pi}} \int^{\infty}_{-\infty} \dd{\phi} e^{- \frac{1}{2} \phi^{2} - \frac{\lambda}{4} \phi^4},
\end{equation}
\noindent counts (vacuum) Feynman diagrams. Indeed, we can expand in powers of \(\lambda\) using ``Wick's Theorem" to write a Feynman diagram expansion. In this one dimensional problem, the ``propagator" is just \(1\); there are no momentum integrals to do. So every diagram at order \(k\) gives simply \(\frac{\qty(-1)^{k}}{k!} \qty(\frac{\lambda}{4})^{k}\). Performing the expansion, order by order, gives:

\begin{equation}
\func{Z}{\lambda} = \sum^{\infty}_{k} z_{k} \qty(\frac{\lambda}{4})^{k},
\end{equation}

\noindent where

\begin{equation}
z_{k} = \frac{1}{\sqrt{\pi}} \frac{\qty(-1)^{k}}{k! 2^{2k}} \func{\Gamma}{2k  + \frac{1}{2}}.
\end{equation}
So we see factorial growth of the number of diagrams, and that the perturbation expansion is an asymptotic expansion, reliable only for \(k \lesssim 1 / \lambda\).  The original integral is, of course, finite for all \(\lambda : \Re(\lambda) > 0\).  The result can be written as a modified Bessel function (See Appendix \ref{app:diagramcounting}); it is easy to check numerically that the series gives good agreement with the exact result if one only includes terms up to \(k\) somewhat smaller than \(1/\lambda\).

For scattering in quantum field theory, we are interested in (connected) \(N\)-point functions. Correspondingly, we can start with the study of the simple one dimensional (Euclidean) integral:

\begin{equation}
\begin{split}
\func{Z}{N,\lambda} =& \frac{1}{\sqrt{2\pi}} \int^{\infty}_{-\infty} \dd{\phi} \phi^{N} e^{- \frac{1}{2} \phi^{2} - \frac{\lambda}{4} \phi^{4}},\\[2mm]
\sim& \frac{1}{\sqrt{\pi}} \sum^{\infty}_{k = 0} \frac{\qty(-1)^{k}}{k!} 2^{-2k - \frac{N}{2}} \func{\Gamma}{\frac{4 k + N + 1}{2}} \qty(\frac{\lambda}{4})^{k}, \quad \qty(\lambda \to 0).
\label{znlambda}
\end{split}
\end{equation}

\noindent Provided \(N \lesssim 1 / \lambda\), such that the asymptotic series reasonably approximates the function, we see that \(\func{Z}{N,\lambda}\) exhibits factorial growth in \(N\),

\begin{equation}
\func{Z}{N,\lambda} \sim \frac{1}{\sqrt{\pi}} 2^{- \frac{N}{2}} \func{\Gamma}{\frac{N+1}{2}}, \quad \qty(\lambda \to 0).
\end{equation}

\noindent The leading connected diagrams occur at order \(k = \qty(N-2)/2\). Correspondingly, we expect a contribution to \(Z\):

\begin{equation}
\begin{split}
\func{Z}{N,\lambda} \sim& - \frac{4}{\sqrt{\pi}} \frac{\qty(-1)^{N/2}}{\qty(\frac{N-2}{2})!} 2^{- 3N/2} \func{\Gamma}{\frac{3 N - 3}{2}} \qty(\frac{\lambda}{4})^{\qty(N-2)/2}, \quad \qty(\lambda \to 0)\\[2mm]
\sim& \qty(-1)^{N/2 + 1} \frac{2^{\qty(7-5N)/2}}{9 \sqrt{\pi}} N^{-3/2} e^{N \ln(N) - N} \qty(\frac{\lambda}{4})^{\qty(N-2)/2}, \quad \qty(\lambda \to 0, N \to \infty)
\end{split}
\end{equation}

\noindent So for this problem, perturbation theory is valid for \(N < 1 / \sqrt{\lambda}\). We will explain shortly, and in an appendix, why this counting, which includes both connected and disconnected diagrams, gives the correct \(N!\) dependence of the amplitudes in this limit.

For \(\lambda N \gg 1/4\) the behavior is different, though the integral still exhibits factorial growth in \(N\) for fixed \(\lambda\). The maximum of the integrand is located at

\begin{equation}
\phi^{4}_{c} + \frac{1}{\lambda} \phi^{2}_{c} - \frac{N}{\lambda} = 0
\end{equation}

\noindent For \(\lambda N\) very large the location of the maximum is shifted substantially away from the origin, and the integrand is dominated by  \(\phi^{4}_{c} \sim  N / \lambda\).

\begin{equation}
\begin{split}
\func{Z}{N, \lambda} \sim& \frac{2^{\qty(N+1)/2}}{N^{1/4} \lambda^{\qty(N+1)/4}} \exp(\frac{N}{4} \ln(\frac{N}{4}) - \frac{N}{4} - \frac{1}{2} \sqrt{\frac{N}{\lambda}} + \frac{5}{16 \lambda}), \quad \qty(N \to \infty).
\label{znestimate}
\end{split}
\end{equation}

\noindent It is easy to check these statements numerically.

So we have learned that, for \(N \gg 1 / \lambda\), \(\func{Z}{N,\lambda}\) exhibits factorial growth, but much slower than the factorial growth of perturbation theory. In subsequent sections, we will argue that the behavior of scattering amplitudes at large \(N\) is similar: they exhibit factorial growth, but \textit{slower} than that of perturbation theory.

\section{From an Ordinary Integral to the Path Integral}
\label{sec:fieldtheorypathintegral}
 
\begin{figure}[ht]
	\centering
	\begin{tikzpicture}[x={(0:1cm)},y={(90:1cm)}]
		\begin{scope}
			\clip (-4,-4) rectangle (4,4);

      \draw[help lines] (-4,-4) grid (4,4);
      \path (2.5,0) arc (0:-11.25:2.5) coordinate[pos=0.5] (anglelabel);
      \node[rectangle, fill=white, text=white, right] at (anglelabel) {\( \arg(\lambda) / 4 \)};

			\node[rectangle] (phi) at (3.5,3.55) {\( \phi \)};
			\draw (phi.north west) -- (phi.south west) -- (phi.south east);

			\filldraw[draw=gray!80, fill=gray!50, fill opacity=0.5] (0,0) -- ++ (11.25:5.657) arc (11.25:-33.75:5.657) -- cycle;
			\filldraw[draw=gray!80, fill=gray!50, fill opacity=0.5] (0,0) -- ++ (101.25:5.657) arc (101.25:56.25:5.657) -- cycle;
			\filldraw[draw=gray!80, fill=gray!50, fill opacity=0.5] (0,0) -- ++ (191.25:5.657) arc (191.25:146.25:5.657) -- cycle;
			\filldraw[draw=gray!80, fill=gray!50, fill opacity=0.5] (0,0) -- ++ (281.25:5.657) arc (281.25:236.25:5.657) -- cycle;

			\draw[thick, latex-latex] (-4,0) -- (4,0);
			\draw[thick, latex-latex] (0,-4) -- (0,4);
			\draw[dashed] (0,0) -- ++(-11.25:5.657);
			\draw[dashed] (0,0) -- ++(78.75:5.657);
			\draw[dashed] (0,0) -- ++(168.75:5.657);
			\draw[dashed] (0,0) -- ++(258.75:5.657);

      \draw (2.5,0) arc (0:-11.25:2.5) node[pos=0.5, right] {\footnotesize \( \arg(\lambda) / 4 \)};
			\path (0,0) -- ++(78.75:2.5) coordinate (a);
			\draw (a) arc (78.75:56.25:2.5) node[pos=0.5, above right] {\footnotesize \( \frac{\pi}{8} \)};
		\end{scope}
	\end{tikzpicture}
	\caption{Regions of convergence of the integral in \eqref{znlambda}. Contour integrals beginning and ending in the gray shaded regions converge. The Euclidean version of the integral is related to the Minkowski version by \(\arg(\phi) = \frac{\pi}{4}\), \(\arg(\lambda) = -\frac{\pi}{2}\).}
	\label{fig:complexphiplane}
\end{figure}
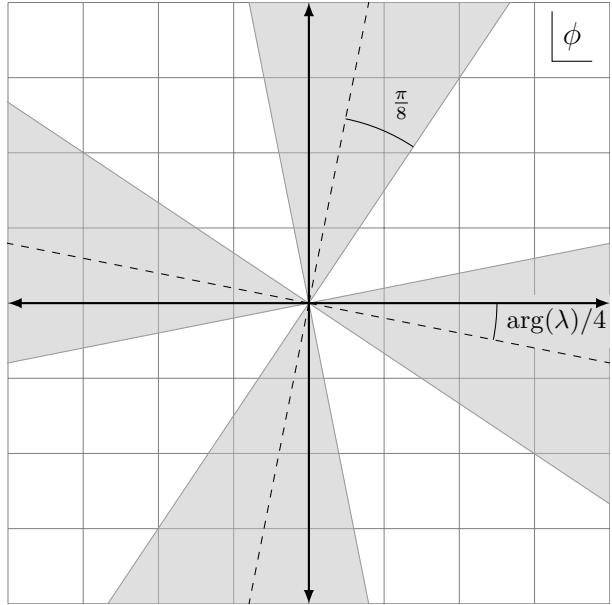

As a model for scattering amplitudes in field theory, the integral \(\func{Z}{N,\lambda}\) is instructive, but has several limitations. The first is a relatively trivial one: the exponential should be a phase (the argument should be purely imaginary). This does not affect our large \(N\) estimate. Analytically continuing the integral \eqref{znlambda} to the complex \(\phi\) plane we find it is absolutely convergent in four wedges, which are rotated by the phase of the coupling (Figure \ref{fig:complexphiplane}).

So the integral defines a function of \(\lambda\) analytic in a wedge. The estimate above holds perfectly well if we simply rotate the variable \(\phi\) by a constant phase angle of \(\frac{\pi}{4}\) and simultaneously rotate the coupling by \(- \frac{\pi}{2}\).

Another limitation is that the integral \eqref{znlambda}, in addition to generating connected diagrams, also generates vacuum and disconnected diagrams. However, the rough factorial growth of the generating function \(\func{Z}{j}\) and \(\func{W}{j} \coloneqq \ln(\func{Z}{j} / \func{Z}{0})\) is the same. This is readily proven by contradiction. Moving over to the full field theory, the generating functional \(\Func{Z}{J}\) is described by a convergent series in inverse powers \(N\). As a result, so is \(\Func{W}{J}\). If terms in the expansion of \(W\) grow more rapidly with \(N\) than those of \(Z\) then, in fact, the growth of \(W\) must be faster than \(Z\); similarly if terms in \(W\) grow more slowly.

This last issue will not be so severe in \(3+1\) dimensions (as opposed to \(0\)) when we work with wave packets, each of some average momentum. Due to momentum conservation, in processes with \(2 \to N\) particles, the associated disconnected contributions to the Green's functions will largely vanish. So it is enough to divide out the vacuum diagrams.
 
The path integral is certainly far more complicated than the ordinary integral, and the question is: to what degree is the behavior of this integral an indicator of what happens in field theory. Various approaches have been offered as solutions to this problem. We will argue that a particularly simple one is to use the LSZ formula, with normalizable wave packets. As a result, the effects of large \(N\) are important only in a small space-time region where the wave packets overlap. One might then expect that, limited to this region, the path integral \textit{would} be much like the ordinary integral. Translating the results for the ordinary integral to the behavior of the field theory path integral is the subject of the next few subsections.

\subsection{Review of Non-Relativistic Scattering in Terms of Wave Packets}
\label{sec:wavepacketscattering}

Our analysis of many particle scattering will rely on a wave packet approach. In particular, it assumes that we can think of the scattering amplitude in terms of normalizable initial and final wave packets labeled by a central momentum \(\vec{p}_{i}\) and \(\vec{p}_{f}\) respectively. It is helpful to review the problem of wave packet scattering in ordinary quantum mechanics. We first consider how to pass between the scattering amplitude in the plane wave basis to a basis of normalizable states\cite{taylor}. Consider, in particular, the scattering amplitude of a particle in a potential, \(\func{V}{\vec{x}}\), with initial and final wave packets, \(\func{\psi_{i}}{t, \vec{x}}\) and \(\func{\psi_{f}}{t, \vec{x}}\). We write the Fourier transform of these wave packets as:

\begin{equation}
\begin{split}
\func{\psi_{i}}{t, \vec{x}} =& \int \frac{\dd[3]{p}}{\qty(2 \pi)^{3}} \func{\tilde{\psi}_{i}}{\vec{p}} e^{- i p \cdot x},\\[2mm]
\func{\psi_{f}}{t, \vec{x}} =& \int \frac{\dd[3]{k}}{\qty(2 \pi)^{3}} \func{\tilde{\psi}_{f}}{\vec{k}} e^{- i k \cdot x}.
\end{split}
\end{equation}

\noindent where \(k^{0} \equiv \func{E}{\vec{k}}\) and \(p^{0} \equiv \func{E}{\vec{p}}\) are the (on-shell) energy. We require the momentum space distribution of scattering states to be narrowly centered on the average momentum, a reasonable fulfillment of this requirement is to approximate the momentum distributions by very narrow Gaussians:

\begin{equation}
\begin{split}
\func{\tilde{\psi}_{i}}{\vec{p}} =& \qty(2 \pi \sigma^{2})^{-3/2} e^{- \frac{\qty( \vec{p} - \vec{p}_{i})^{2}}{2 \sigma^{2}}} e^{i \vec{p} \cdot \vec{x_0^i}},\\[2mm]
\func{\tilde{\psi}_{f}}{\vec{k}} =& \qty(2 \pi \sigma^{2})^{-3/2} e^{- \frac{\qty( \vec{k} - \vec{k}_{f})^{2}}{2 \sigma^{2}}}e^{i \vec{k} \cdot \vec{x_0^f}}.
\end{split}
\end{equation}

\noindent We have taken the spread in the momentum space wave packets to be identical for simplicity. At non-relativistic energies the coordinate space center of these wave packets propagate as

\begin{equation}
\begin{split}
\expval{\vec{x}}_{i} \sim \vec{x_0^i} +& \frac{\vec{p}_{i}}{m} t , \quad \qty(\sigma \to 0),\\[2mm]
\expval{\vec{x}}_{f} \sim \vec{x_0^f}+ & \frac{\vec{p}_{f}}{m} t, \quad \qty(\sigma \to 0),
\end{split}
\end{equation}

\noindent and have a widths that increase with time. We will assume, again for simplicity, that this spreading can
be neglected during the scattering process, so \(\Delta x \approx \frac{1}{\sigma}\). To obtain an appreciable amplitude it is necessary that the trajectories of the two waves coincide during some time interval of order \(m / \sigma\).

We'll confine our attention to the Born approximation for the scattering amplitude and work in the interaction picture. We first consider as a basis of scattering states (non-normalizable) plane  waves. We can write the \(S\)-matrix in terms of the interaction picture time-development operator, \(\func{U}{t_{1}, t_{2}}\) as (taking the limit \(T \to \infty\)):

\begin{equation}
\bra{\vec{p}_{f}} \func{U}{T/2, -T/2} \ket{\vec{p}_{i}} \coloneqq \bra{\vec{p}_{f}} S \ket{\vec{p}_{i}}
\end{equation}

\noindent where

\begin{equation}
\bra{\vec{p}_{f}} S \ket{\vec{p}_{i}} = \bra{\vec{p}_{f}}\ket{\vec{p}_{i}} + \qty(2 \pi i) \ddelta{E_{f} - E_{i}} \bra{\vec{p}_{f}} H^\prime \ket{\vec{p}_{i}}.
\end{equation}

\noindent The second term is the \(T\) matrix. In terms of the plane wave basis, the scattering amplitude for our wave packets, \(\psi_{i}\), \(\psi_{f}\) is

\begin{equation} \label{eq:bornamp}
\cA_{i \to f} = \int \frac{\dd[3]{p_{i}}}{\qty(2 \pi)^{3}} \frac{\dd[3]{p_{f}}}{\qty(2 \pi)^{3}} \func{\tilde{\psi}^{*}_{f}}{\vec{p}_{f}} \bra{\vec{p}_{f}} S \ket{\vec{p}_{i}} \func{\tilde{\psi}_{i}}{\vec{p}_{i}}.
\end{equation}

\noindent Calculating the amplitude to first order in perturbation theory in terms of normalizable states in the interaction picture, we can understand the expression in Equation \eqref{eq:bornamp} in another way. In general, far away from the forward direction, the amplitude that an initial state \(\ket{\Psi_{i}}\) time evolves into a state \(\ket{\Psi_{f}}\) is:

\begin{equation}
\cA_{i \to f} = \int^{T/2}_{-T/2} \bra{\Psi_{f}} e^{i H_{0} t} H^{\prime} e^{-i H_{0} t} \ket{\Psi_{i}} \dd{t}. 
\end{equation}

\noindent We can rewrite this in terms of Schr\"{o}dinger picture states:

\begin{equation}
\cA_{i \to f} = \int \bra{\func{\Psi_{f}}{t}} H^{\prime} \ket{\func{\Psi_{i}}{t}} \dd{t}.
\end{equation}

\noindent So for our scattering problem,

\begin{equation}
\begin{split}
\cA_{i \to f} =& \int^{T/2}_{-T/2} \dd{t} \int \frac{\dd[3]{p_{f}}}{\qty(2 \pi)^{3}} \frac{\dd[3]{p_{i}}}{\qty(2 \pi)^{3}} \func{\tilde{\psi}^{*}_{f}}{t, \vec{p}_{f}} \bra{\vec{p}_{f}} H^{\prime} \ket{\vec{p}_i} \func{\tilde{\psi}_{i}}{t, \vec{p}_{i}}\\[2mm]
=& \int^{T/2}_{-T/2} \dd{t} \int \frac{\dd[3]{p_{f}}}{\qty(2 \pi)^{3}} \frac{\dd[3]{p_{i}}}{\qty(2 \pi)^{3}} \func{\tilde{\psi}^{*}_{f}}{\vec{p}_{f}} e^{-i \qty(E_{i} - E_{f}) t} \bra{\vec{p}_{f}} H^{\prime} \ket{\vec{p}_{i}} \func{\tilde{\psi}_{i}}{\vec{p}_{i}}\\[2mm]
=& 2 \pi \int \frac{\dd[3]{p_{f}}}{\qty(2 \pi)^{3}} \frac{\dd[3]{p_{i}}}{\qty(2 \pi)^{3}} \func{\tilde{\psi}^{*}_{f}}{\vec{p}_{f}} \ddelta{E_{f} - E_{i}} \bra{\vec{p_{f}}} H^{\prime} \ket{\vec{p_{i}}} \func{\tilde{\psi}_{i}}{\vec{p}_{i}}
\end{split}
\end{equation}

The second line is equivalent to the coordinate space expression:
\begin{equation}
\cA_{i \to f} = \int^{T/2}_{-T/2} \dd{t} \int \dd[3]{x}  \func{\psi^{*}_f}{t, \vec{x}} \func{H^{\prime}}{\vec{x}} \func{\psi_{i}}{t, \vec{x}}.
\end{equation} 
 
If \(H^\prime\) is short range, then this last expression has support only when the wave functions
overlap in a space-time region with size less than or order the range of the potential. Note that in order to obtain a cross section from this, one must take \(\abs{\cA_{i \to f}}^2\), and integrate over \(\qty(\dd[2]{x_{i}})_\bot \equiv \dd[2]{b}\) weighted by the flux (this assumes, for simplicity, identical wave packets up to translations). One also integrates over and \(\dd[3]{x_{f}}\), for fixed \(\vec{b}\), \(\vec{p}_{i}\), and \(\vec{p}_{f}\). The result, again, is appreciable only for a range of \(\vec{x}_{f}\) where the wave packet points back to the interaction point. In this range, one can replace \(\dd[3]{x_{f}} = r^{2} \dd{r} \dd{\Omega}\), allowing construction of the differential cross section. This will, in fact, give a contribution proportional to the ``cross section" of the target, the fractional region over which the integrand is substantial. For a given \(\vec{p}_{f}\), this fixes \(\vec{x}_{f}\) to lie in a small region, \(\abs{\vec{x}_{f}} < \mu^{-1}\), the range of the potential.
 
This generalizes immediately to multichannel problems with, for instance, one particle initially impinging on a target (say a high \textit{Z} atom), and many particles emerging. Again, one has an overlap of the Schr\"{o}dinger wave function for the initial state, evolved with the free particle Hamiltonian to some time \(t\), and the final state particles, evolved back to time \(t\). To obtain an appreciable result, there must be a time \(t\) where all of the wave functions coincide within a space-time region of order the range of the potential. Again, the (differential) cross section is obtained by integrating over \(\dd[2]{x_{i}}\) and dividing by the incoming flux.

If one formulates the amplitude in terms of the path integral, the integration \(\int \qty[\dd{x}]\) is only significantly modified from the free particle result for the narrow range of variables \(\func{x}{t}\) corresponding to the time interval where the wave packets overlap.
 
\subsection{Scattering of Wave Packets using the LSZ Formula and the Path Integral}
\label{sec:lsz}

We want to consider processes with many particles in the initial and/or final state. Our goal is to determine the growth of the scattering amplitude for \(N \gg 1/\lambda\).

The LSZ formula for scattering casts the problem of scattering in terms of Green's functions, so it is a natural setting in which to apply path integral methods. There is some discussion of wave packet scattering in this framework in textbooks, e.g.~\cite{peskinschroeder}. We will review this here, from a perspective close to the non-relativistic problem which we have described in the previous section, and which will be useful for the questions we are studying here.

Let's consider \(\phi^{4}\) theory. In the LSZ formula, for a scattering process with \(M\) particles in the initial state and \(N\) particles in the final state, we are instructed to evaluate the quantity:

\begin{equation}
\prod^{M+N}_{i=1} \left[ \int \dd[4]{x_{i}} \func{f_{i}}{x_{i}} \qty(\partial^{2}_{i} + m^2) \right] \expval{ \func{\phi}{x_{1}} \ldots \func{\phi}{x_{N+M}}},
\label{lszformula}
\end{equation}

\noindent where the functions \(f_{i}\) satisfy the free Klein-Gordan equation. We would like to treat this expression in the path integral, studying possible modifications associated with the large number of particles in the initial and final states, \(M+N\). The inverse propagators in Equation \eqref{lszformula} make this somewhat awkward, particularly when we attempt a large \(N\), semiclassical treatment. One approach is to consider \textit{classes} of diagrams where one contracts the external fields with fields in vertices, and evaluates the remaining Green's function. One then has to sum over the different classes.

An alternative is also useful. One traditional derivation of the LSZ formula starts with initial and final states at times \(t = \pm T\), and evaluates:

\begin{multline}
\cM = \prod^{M}_{i = 1} \left[ \int \dd[3]{x_{i}} \func{\psi_{i}}{\vec{x}_{i}} \right] \prod^{N}_{j = 1} \left[ \int \dd[3]{y_{j}} \func{\xi_{j}}{\vec{y}_{j}} \right]\\[2mm]
\times \expval{\func{\phi}{-T, \vec{x}_{1}} \ldots \func{\phi}{-T, \vec{x}_{M}} \func{\phi}{T, \vec{y}_{1}} \ldots \func{\phi}{T, \vec{y}_{N}}}.
\end{multline}

\noindent Here \(\psi_{i}\), \(\xi_{j}\) are initial and final state wave functions at times \(\pm T\). They are taken to be normalizable and non-overlapping. The correlation function can be studied perturbatively. A non-perturbative approach could involve construction of a one-particle irreducible effective action at \(\qty(N+M)\)\textsuperscript{th} order in \(\phi\). If this interaction, \(\Gamma_{M \to N}\), is local the structure of the resulting path integral has many features in common with the toy integral. In particular, only the integration over a small local region of space-time is modified by large \(N\).

\section{Non-Perturbative Behavior of the Amplitude: \texorpdfstring{\(N \rightarrow N\)}{N to N}}
\label{sec:nonperturbativenton}

We have argued that we can reduce the problem of computing the scattering rate for large numbers of particles to a problem of summing over a finite set of correlation functions which can be evaluated using the path integral. We might hope that large \(N\) might facilitate a non-perturbative evaluation, as in the case of the ordinary integrals we studied in previous sections. The use of wave packets of finite extent enhances the fields in a small region, mimicking some features of the ordinary integral at large \(N\).

Consider, first, the case of \(N \to N\) scattering. We saw that, in perturbation theory, there is a region of the phase space integral for which one class of diagrams is enhanced dynamically. There were of order \(\qty(N!)^{2}\) diagrams in this subset (coming from \(N!\) rearrangements of the initial state particles and \(N!\) rearrangements of the final state particles) and a factor of \(1/N!\) due to the number of vertices, but the corresponding \(\qty(N!)^{2}\) enhancement in the squared amplitude was compensated by the Bose statistics factors associated with the initial and final states. The number of diagrams without any kinematic restriction grows as \(\qty(2N)!\), but arguably the dynamical dependence of the typical diagram on \(N\) compensates this growth. On the other hand, even for the class of diagrams where the growth is not factorial, we have not yet described a systematic approach to the computation of the amplitudes, and any claim for the general behavior is conjectural. Perturbatively, the kinematically enhanced region arises when pairs of initial and final momenta are nearly the same, \(\vec{p}_{i} = \vec{k}_{i} + \delta \vec{p}_{i}\). So calling \(f_{i}\) the initial state wave functions, with momenta centered around \(\vec{p}_i^0\) and width \(\Delta p\), and similarly denoting the final state wave functions and mean momenta by \(g_i\), \(\vec{k}_{i}\), we are interested in the set of correlation functions, of which one example (dropping terms of order \(1\)) is:

\begin{equation}
\cA = \prod^{N}_{i=1} \sum_{{\rm perm}~g_{i}} \int \dd[4]{x_{i}} \expval{\func[2]{\phi}{x_{1}} \ldots \func[2]{\phi}{x_{N}}} \func{f_{1}}{x_{1}} \func{g^{*}_{1}}{x_{1}} \ldots \func{f_{N}}{x_{N}} \func{g^{*}_{N}}{x_{N}}.
\end{equation}

\noindent This integral is much like our one dimensional toy example in two ways. First, writing this as a path integral and exponentiating the fields appearing in the Green's function:

\begin{equation}
\cA = \int [\dd{\phi}] e^{i \int \dd[4]{x} \left\{ \frac{1}{2} \left(\partial_{\mu} \phi \right)^2 - \frac{m^{2}}{2} \phi^{2} - \frac{\lambda}{4} \phi^{4} \right\}  + \sum^{N}_{i=1} \ln[\int \dd[4]{x} \func{f_{i}}{x} \func{g^{*}_{i}}{x} \func[2]{\phi}{x}]}. 
\end{equation}

\noindent The logarithmic term in the exponent is enhanced when all of the points nearly coincide. This enhancement is largest if pairs of the \(f_{i}\)'s and \(g_{i}\)'s are nearly the same, in which case the exponent is of order \(N\). This is similar to the configurations of parallel momenta we discussed in perturbation theory. For random permutations, we might expect the exponent to be of order \(\sqrt{N}\). If we assume that the path integral is dominated by a particular classical configuration, \(\func{\phi_{\text{cl}}}{x}\), we have:

\begin{equation}
\qty(\partial^2 + m^2) \func{\phi_{\text{cl}}}{x} = \lambda \func[3]{\phi_{\text{cl}}}{x} + 2 \sum^{N}_{i=1} \frac{ \func{f_{i}}{x} \func{g^{*}_{i}}{x} \func{\phi_{\text{cl}}}{x}}{\int \dd[4]{y} \func{f_{i}}{y} \func{g^{*}_{i}}{y} \func[2]{\phi_{\text{cl}}}{y}}
\end{equation}

\noindent For parallel or nearly parallel pairs of momenta and identical or nearly identical initial and final state wave packets, the second term on the right hand side is of order \(N \phi^{-1}_{\text{cl}}\). The right hand side of the equation can be made to vanish if \(\phi_{\text{cl}} \sim N^{1/4}\); the left hand side of the equation is then of order \(N^{1/4}\), i.e. suppressed by \(N^{-1/2}\) relative to the separate terms on the right hand side. In this case, the dominant term in the classical action is of order \(N \ln(N) / 2\), giving an amplitude growing as \(\qty(N/2)!\), in contrast to the tree level growth of \(N!\) found using perturbation theory. Alternatively, we can literally view the path integral as like our ordinary integral, thinking of \(\func{\phi}{0}\) as a single variable. This yields the same \(N!\) dependence as above.

For non-parallel momentum pairs, we might expect a suppression, as in the perturbative case. From this latter point of view, the contractions of the external fields where the pairs of fields have \textit{different} momenta involve integrations over \textit{different} variables. In this case, we might not expect the modification due to the large value of \(N\) to be captured by the model integral or the semiclassical solution. Indeed, we would not expect appreciable modifications from the perturbative result. If we attempt an analysis of the sort we did for parallel momenta in perturbation theory for non-parallel momenta, assuming that the terms in the exponent add with random phases, we would find a contribution for each contraction behaving  as \(\qty(N/8)!\). The \(N!\) contributions then would also add with random phases, yielding a contribution to the amplitude behaving as \(\qty(3N/8)!\). Squaring and dividing by the Bose statistics factors would yield a slightly larger contribution to the cross section, but still falling with \(N\) for large \(N\).

We should ask: to what extent is this analysis systematic? For the subset of contributions where the momenta are paired, we can give a rough argument that corrections to the leading approximation are down by powers of \(N\). We distinguish two types of corrections: corrections to the classical solution and ``loop" corrections to the amplitude. Substituting back in the original equation and writing

\begin{equation}
\phi_{\text{cl}} = \phi_{\text{cl}}^{0} + \delta \phi_{\text{cl}} + \delta \phi_{\text{quant}},
\end{equation}

\noindent where \(\delta \phi_{\text{cl}} \sim N^{-1/4}\), i.e. the expansion of the classical solution about the leading result appears to be an expansion in powers of \(N^{-1/2}\). Quantum (loop) corrections have a \(\delta \phi_{\text{quant}}\) propagator proportional to \(N^{-1/2}\), and three point vertices of order \(N^{1/4}\), so loop corrections appear to scale with \(N^{-1/2}\) as well. So, while we will not investigate this further here, it appears that for these processes there is a systematic \(1/N\) expansion. Establishing this requires demonstrating that classes of contributions with different contractions of the external fields are indeed suppressed.

\section{Non-Perturbative Analysis: \texorpdfstring{\(2 \to N\)}{2 to N} Scattering}
\label{sec:nonperturbative2ton}

For \(2 \to N\) processes we have not found a systematic \(1/N\) expansion. If we proceed as we did for \(N \to N\) scattering we encounter a functional integral whose integrand involves multiple integration variables (roughly \(\func{\phi}{nm,\vec 0}\) and \(\func{\phi}{rm,\vec0}\)), whose coupling is complicated. However, similar considerations suggest that the non-perturbative growth of the amplitude is far slower than the perturbative one. For instance, suppose we have computed an effective action for \(2 \to N\) particles,

\begin{equation}
\cL_{I} = \frac{\Gamma_{2 \to N}}{N!} \phi^{N}.
\end{equation}

\noindent We expect that if, in the center of mass frame, the typical spatial momenta is of order \(\abs{\vec{p}} \ll m\), then \(\Gamma_{2 \to N}\) is approximately independent of \(p\). It is convenient to avoid the factors of inverse propagators, so with our \(N\) final state particles with wave functions \(\func{\psi_{i}}{\vec{x}}\) in the Schr\"{o}dinger picture at some large time \(T\), we need to study:

\begin{equation}
\cM_{2 \to N} = \frac{\Gamma_{2 \to N}}{N!} \prod^{N}_{i} \left[ \int \dd[3]{x_{i}} \func{\psi_{i}}{\vec{x}_{i}} \right] \int \dd[4]{z} \expval{\func[N]{\phi}{z} \func{\phi}{T, \vec{x}_{1}} \ldots \func{\phi}{T, \vec{x}_{N}}}.
\label{modifiedlsz}
\end{equation}

\noindent Note that the wave functions have support only when \(\vec{x}_{1}, \ldots, \vec{x}_{N}\) are well separated at time \(t = T\), and we have explicitly implemented the assumption of locality.

To determine the dependence of the scattering amplitude on \(N\) we will proceed in two steps. First, Then we will determine the \(N\) dependence of the Green's function appearing in the expression for \(\cM_{2 \to N}\) in Equation \eqref{modifiedlsz}.

Consider, first, the problem in perturbation theory. We can write an iterative relation between \(\Gamma_{2 \to N}\) and \(\Gamma_{2 \to N/3}\); these correspond to terms in the effective Lagrangian:

\begin{align}
\cL_{\text{eff}} =& \Gamma_{2 \to N} \frac{\func[N]{\tilde{\phi}}{m, \vec{0}}}{N!} & \cL_{\text{eff}} =& \Gamma_{2 \to N/3} \frac{\func[N/3]{\tilde{\phi}}{3 m,\vec{0}}}{\qty(N/3)!}.
\end{align}

\noindent where \(\tilde{\phi}\) is the momentum space field and \(m\) is the scalar mass. For general \(N\) we can compute the \(N\) point Green's function, either starting with \(\Gamma_{2 \to N}\) or with \(\Gamma_{2 \to N/3}\), and expand the path integral to order \(N/3\) in the interaction:

\begin{multline}
\frac{\Gamma_{2 \to N}}{N!} \expval{\func[N]{\tilde{\phi}}{m, \vec{0}} \func[N]{\phi}{m, \vec{0}}} =\\[2mm]
\frac{\lambda^{N/3}}{\qty(N/3)!} \frac{\Gamma_{2 \to N/3}}{\qty(N / 3)!} \expval{\func[N/3]{\tilde{\phi}}{3 m, \vec{0}} \func[N]{\tilde{\phi}}{m, \vec{0}} \func[N]{\tilde{\phi}}{m, \vec{0}} \func[N/3]{\tilde{\phi}}{3 m, \vec{0}}}.
\end{multline}

\noindent The correlation functions can be evaluated just as for our one dimensional
integrals.   For \(N \ll 1 / \lambda\) gives

\begin{equation}
\Gamma_{2 \to N} = \frac{N!}{\qty(N/3)!} \Gamma_{2 \to N/3}.
\end{equation}

\noindent This is solved by

\begin{equation}
\Gamma_{2 \to N} = C N!,
\end{equation}

\noindent where \(C\) is a constant, as expected from perturbation theory (the constant can be determined by matching to the perturbative result). If we consider the limit \(N \gg 1 / \lambda\), proceeding as in Equation \eqref{znestimate} we obtain the recursion relation:

\begin{equation}
\begin{split}
\frac{\qty(N / 2)!}{N!} \Gamma_{2 \to N} =& \qty(N / 2)! \qty(N /6)!  \frac{1}{\qty(N / 3)!^2} \Gamma_{2 \to N/3},\\[2mm]
\Rightarrow \Gamma_{2 \to N} =& C \qty(N/2)! \Gamma_{2 \to N/3}
\end{split}
\end{equation}

\noindent This is solved by

\begin{equation}
\Gamma_{2 \to N} = C \qty(\frac{3 N}{4})!,
\end{equation}

\noindent where, again, \(C\) is some arbitrary constant. With this result we can consider the scattering amplitude using the version of the LSZ formula of Equation \eqref{modifiedlsz}. For the perturbative case we recover the result \(\cM \sim N!\). For the case when \(\lambda N\) is large, we have instead:

\begin{equation} \label{eq:npM2toN}
\begin{split}
\cM \sim& \qty(\frac{N}{2})! \qty(\frac{3N}{4})! \frac{1}{N!}, \quad \qty(N \to \infty),\\[2mm]
\sim& \qty(\frac{N}{4})!, \quad \qty(N \to \infty)
\end{split}
\end{equation}

\noindent The first factor of Equation \eqref{eq:npM2toN} is that which we have just derived for \(\Gamma_{2 \to N}\); the second is from the correlation function of \(N\) fields near the same point, the third is from the definition of the effective action. This result gives a \textit{cross section} which falls off as

\begin{equation}
\sigma_{2 \rightarrow N} \sim \frac{1}{\qty(N/2)!} \quad \qty(N \to \infty)
\end{equation}

This analysis is not systematic. In particular, in deriving our would-be non-perturbative recursion relation for \(\Gamma_{2 \to N}\), we performed an expansion of the exponent of the interaction term in powers of \(\lambda\). Despite the previous concerns we make the following observations:

\begin{enumerate}
  \item  This analysis makes clear that the leading perturbative result is misleading and likely \textit{vastly} overestimates the amplitude for large \(N\).
  \item  Given that we have only considered, in effect, the summation of an infinite
  \textit{subclass} of diagrams, it is likely that we still overestimate the result.
\end{enumerate}

\section{Conclusions} \label{sec:conclusion}

In quantum field theories perturbative expansions of observables around small coupling are almost always divergent asymptotic expansions; if \(\lambda\) is a typical coupling constant, the number of Feynman diagrams at order \(k\) is typically of order \(\qty(2k)!\), and the series approximates the actual theory only for \(k \lesssim 1 / \lambda\).  

For scattering processes involving large numbers of particles there is also factorial growth of the number of diagrams, now with the \textit{number of particles}, \(N\), as well as the order of perturbation theory. This raises two prospects:  First, that perturbation theory is not a reliable tool for computing scattering amplitudes, for sufficiently large \(N\); and second, that the amplitudes might themselves grow quickly with \(N\), endangering unitarity. In this paper we have investigated both of these issues. We focused on two classes of processes: \(N \rightarrow N\) particles scattering, with all particles near threshold, and \(2 \to N\) scattering. We first reviewed the situation in the lowest non-trivial order in perturbation theory. In \(N \to N\) scattering, there is \(N!\) growth in the amplitude; Allowing for Bose statistics factors, this class of contributions to the cross section does not show factorial growth. Without this restriction there are vastly more diagrams, so there is the potential for such growth, even if the vast majority of the diagrams are not kinematically enhanced.  We gave crude arguments that these other diagrams actually have factorial suppression. In the case of \( 2 \to N\) scattering the amplitude \textit{does} grow as \(N!\), so the amplitude-squared as \(\qty(2N)!\). There is a \(1/N!\)  from Bose statistics, so the lowest order contribution to the cross section \textit{does} grow factorially.

In both cases we have noted that, for \(N \gg 1/\lambda\), perturbation theory breaks down. It is not \textit{a priori} clear that any standard non-perturbative tools are available for a systematic computation. To obtain some insight into the non-perturbative problem, we have studied scattering in \(\lambda \phi^4\) theory in a path integral framework. To set up the problem we have considered scattering of normalizable wave packets and worked with the LSZ expression for the scattering amplitude. Because the wave packets are localized, there is a significant modification of the path integral from the free-field form only in a small space-time region where all of the wave packets overlap. We set up the path integral problem in two ways: one more suitable for the \(N \to N\) process, the other more suitable for the \(2 \to N\) process.
 
In the case of \(N \to N\) scattering we argued that the dominant contribution is due to diagrams with pairs of momenta nearly equal. This corresponds to a class of contributions which can be treated semiclassically, with the amplitude growing factorially, but more slowly than in perturbation theory, and corrections which can be computed as a series in \(1/\sqrt{N}\). As a result, the cross sections for large \(N\) are suppressed, and there are no issues with unitarity. We gave some arguments that the approximation is systematic, though we will not claim they are rigorous.

For \(2 \to N\) scattering we reorganized the computation in terms of an effective action for \(2 \to N\). Here our tool was a recursion relation for \(\Gamma_{2 \to N}\). Taking the limit \(N \ll \lambda^{-1}\) reproduced the results from perturbation theory. For \(N \gg \lambda^{-1}\), this relation yielded much slower growth, so that the scattering cross section does not show factorial growth. As we explained, the calculation is not systematic, but it likely \textit{overestimates} the cross section.

We note that the authors of \cite{higgsplosion1,higgsplosion2} have also formulated the problem in terms of classical field evolution\cite{khozereview}. Such a system can be described in the language of coherent states, with large values for the field eigenvalue, corresponding to large occupation numbers. This problem is slightly different than the one we have considered here, where we had many widely separated particles; the coherent state problem would correspond to large numbers of particles in, say, two identical incoming and two identical outgoing states. But in this case, the problem is equivalent to classical evolution. The classical cross section is limited by energy conservation; one can't have the equivalent of factorial growth.
  
What is perhaps interesting in these problems is that there is a regime of quantum field theory for which, even at weak coupling, our standard tools of analysis fail to yield reliable results. We view our work as providing a strategy to explore this domain. It would be desirable to make the \(N \to N\) analysis more solid, and to make further inroads in the \(2 \to N\) problem, perhaps proving rigorous bounds if not providing a systematic approximation procedure.

\vskip 1cm 
\noindent {\bf Acknowledgments:} This work was supported in part by the U.S. Department of Energy grant number DE-FG02-04ER41286.

\printbibliography

\appendix

\section{Diagram Counting} \label{app:diagramcounting}

In the case of \(\lambda \phi^{4}\) theory we have the generating function, in \(0\) dimensions:
\begin{equation} \label{eq:jpartition}
\func{Z}{\lambda, j} \coloneqq \frac{1}{\sqrt{2 \pi}} \int^{\infty}_{- \infty} \dd{\phi} \exp(- \frac{1}{2} \phi^{2} - \frac{\lambda}{4} \phi^{4} + j \phi),
\end{equation}%
where we have set \(m = 1\) for simplicity and included an external source term. Expanded to order \(\lambda^{n}\), the generating function simply counts the number of graphs (defined as the number of Wick contractions) at each order. For general \(\lambda\), this can be studied as an ordinary integral.

The generating function for graphs with \(N\) external legs is then 
\begin{equation} \label{eq:Npartition}
\begin{split}
\dfunc[N]{Z}{\lambda} \coloneqq& \eval{\dv[N]{j} \func{Z}{\lambda, j}}_{j = 0},\\[2mm]
=& \frac{1}{\sqrt{2 \pi}} \int^{\infty}_{- \infty} \dd{\phi} \phi^{N} \exp(- \frac{1}{2} \phi^{2} - \frac{\lambda}{4} \phi^{4}).
\end{split}
\end{equation}%
This integral vanishes for odd \(N\), so we let \(N \equiv 2 n\) so that we may consider only the non zero components. One can expand around small \(\lambda\). After exchanging the order of integration and summation we find:%
\begin{equation} \label{eq:2npartition}
\begin{split}
\dfunc[2n]{Z}{\lambda} \sim& \frac{1}{\sqrt{2 \pi}} \sum^{\infty}_{k = 0} \frac{\left( - 1 \right)^{k} 2^{2 k + n + 1/2} \func{\Gamma}{2 k + n + \frac{1}{2}}}{k!} \left( \frac{\lambda}{4} \right)^{k}, \quad \left( \lambda \to 0 \right),\\[2mm]
\sim& \sum^{\infty}_{k = 0} \frac{\left( - 1 \right)^{k} \left( 4 k + 2 n - 1 \right)!!}{k!} \left( \frac{\lambda}{4} \right)^{k}, \quad \left( \lambda \to 0 \right).
\end{split}
\end{equation}%
The total number of graphs at each order \(k\) is the absolute value of each coefficient after factoring out the \(k!\) in the denominator:%
\begin{equation} \label{eq:totalgraphsallorders}
\eta_{2n, k} = \left( 4 k + 2 n - 1 \right)!!.
\end{equation}

The integral \eqref{eq:Npartition} can be expressed as a sum of modified Bessel functions. The rapid growth of the coefficients means the series is asymptotic, with a radius of convergence of 0. We are interested, in particular, in the subset of these graphs which are fully connected. One may wonder if removing the vacuum and disconnected graphs might reduce the factorial growth of the coefficients in \eqref{eq:2npartition}. The connected diagram generating function is related to the full generating function via%
\begin{equation} \label{eq:connectedpartition}
\func{W}{\lambda, j} \coloneqq - \ln(\frac{\func{Z}{\lambda, j}}{\func{Z}{\lambda, 0}}).
\end{equation}%
To determine the behavior of \(W\) at leading order we employ the Schwinger-Dyson equations (reintroducing \(\hbar\) as a loop counting parameter).%
\begin{equation} \label{eq:schwingerdyson}
\left[ - \hbar \dv{j} - \lambda \hbar^{3} \dv[3]{j} + j \right] \func{Z}{\lambda, j} = 0.
\end{equation}%
Substituting \eqref{eq:connectedpartition} into \eqref{eq:schwingerdyson}:%
\begin{equation} \label{eq:schwingerdysonconnected}
\dv{\func{W}{\lambda, j}}{j} + \lambda \left( \dv{\func{W}{\lambda, j}}{j} \right)^{3} - 3 \lambda \hbar \dv{\func{W}{\lambda, j}}{j} \dv[2]{\func{W}{\lambda, j}}{j} + \lambda \hbar^{2} \dv[3]{\func{W}{\lambda, j}}{j} + j = 0.
\end{equation}%
Taking \(\hbar \to 0\) in \eqref{eq:schwingerdysonconnected} results in the loop expansion of Feynman diagrams. The leading order (or tree-level) term, \(W_{0}\), is the solution of a simple cubic equation. The real root is easily solved for using Cardano's formula for depressed cubics. One then integrates the cubic equation and applies the boundary condition \(\func{W}{0} = 0\). Expanding the solution for \(j \to 0\) yields an asymptotic series whose coefficients give the number of connected graphs at tree-level with \(2n\) external legs:%
\begin{equation}
\func{W_{0}}{\lambda, j} = \sum^{\infty}_{n = 0} \frac{\left( - 1 \right)^{n} 4^{n-1} \gammafunc{3 n - 2}}{\gammafunc{2n+1} \gammafunc{n}} \left( \frac{\lambda}{4} \right)^{n - 1} j^{2 n} - \frac{1}{6 \lambda}.
\end{equation}%
We can compare this with the corresponding behavior of \(\dfunc[2n]{Z}{\lambda}\). The tree-level connected graphs arise at order
\begin{equation} \label{eq:treeleveldef}
k = n - 1.
\end{equation}%
Calculating the coefficient of \(\dfunc[2n]{Z}{\lambda}\) at this order by inserting \eqref{eq:treeleveldef} into \eqref{eq:totalgraphsallorders} we find that (reverting to \(N=2n\)),

\begin{equation}
z_{n-1} \sim N! \sim \eval{\dv[N]{W_{0}}{j}}_{j = 0}, \quad \qty(N \to \infty),
\end{equation}
where \(\dfunc[2n]{Z}{\lambda} = \sum z_{k} \qty(\lambda / 4)^{k}\). As we noted in the text, the similarities in the behavior of the full and connected Green's functions are expected for large \(N\).

\end{document}